\documentclass[12pt,preprint]{aastex}
\usepackage{pslatex}
\begin{document}

\title{Acceleration of energetic particles by large-scale compressible
magnetohydrodynamic turbulence}
\author{Benjamin D. G. Chandran }
\email{ benjamin-chandran@uiowa.edu} 
 \affil{Department of Physics \& Astronomy, University of Iowa, Iowa City, IA 52242}
\author{Jason L. Maron}
\affil{Department of Physics \& Astronomy, University of Iowa, Iowa City, IA 52242;\\
Department of Physics \& Astronomy, University of Rochester, Rochester, NY, 14627-0171}

\begin{abstract}
Fast particles diffusing along magnetic field lines in a
turbulent plasma can diffuse through and then return to the same eddy
many times before the eddy is randomized in the turbulent flow.  This
leads to an enhancement of particle acceleration by large-scale
compressible turbulence relative to previous estimates in which
isotropic particle diffusion is assumed.
\end{abstract}
\maketitle

\section{Introduction}

Particle acceleration by magnetohydrodynamic (MHD) turbulence has been
investigated by a number of authors (e.g., Fermi 1949, Hall \& Sturrock 1967,
Wentzel 1968, Kulsrud \& Ferrari 1971, Eichler 1979, Achterberg 1981,
Schlickeiser \& Miller 1998), and has been studied in a number of
astrophysical settings, including the interstellar medium (Ptuskin
1988, Schlickeiser \& Miller 1998), solar flares (LaRosa et~al 1994,
Miller et~al 1996, Miller et~al 1997, Miller 1998, Schlickeiser \&
Miller 1998, Miller 1998), active galactic nuclei (Blackman 1999,
Gruzinov \& Quataert 1999), diffuse intracluster plasma in galaxy
clusters (Eilek \& Weatherall 1999, Weatherall \& Eilek 1999, Dogiel
1999, Brunetti et~al 1999, Ohno et~al 2002, Fujita et~al 2003), and the
lobes of extragalactic radio sources (Eilek \& Henriksen 1984,
Manolakou et~al 1999).

The way in which magnetohydrodynamic (MHD) turbulence accelerates
energetic particles depends on the relative magnitudes of
$\lambda_\parallel$ and $\lambda_{\rm mfp}$, where $\lambda_\parallel$
is the length of a turbulent eddy (or the wave length of a weakly
damped wave) measured along the magnetic field, and $\lambda_{\rm
mfp}$ is the energetic-particle scattering mean free path.  In the
weak-scattering limit, $\lambda_\parallel \ll \lambda_{\rm mfp}$,
 a particle's velocity~$v_\parallel$ along the
background magnetic field is approximately constant as the particle
travels many wave lengths along the magnetic field (assuming that the
amplitudes of the fluctuations are sufficiently small), and the particle
interacts resonantly with a weakly-damped wave when the wave frequency
in the frame of reference moving along the magnetic field at
speed~$v_\parallel$ is an integer multiple of the particle's gyrofrequency.
In the strong-scattering limit, $\lambda_\parallel \gg \lambda_{\rm mfp}$,
a particle travels diffusively across an eddy of length~$\lambda_\parallel$,
and acceleration is non-resonant.

Particle acceleration in the strong-scattering limit is the focus of
this paper.  Ptuskin (1988) studied particle acceleration in the
strong-scattering limit by a power-law spectrum of acoustic or
magnetosonic waves extending from a large scale~$l$ to a smaller
scale~$l_1$, with rms fluctuation velocity~$u_{\rm rms}$ dominated by the waves
at scale~$l$. He assumed that particle diffusion in space is
isotropic with diffusion coefficient~$D$, as opposed to primarily
along field lines, and that some type of small-scale wave distinct
from the acoustic/magnetosonic waves scatters the particles, with $\lambda_{\rm
mfp} \sim D/v \ll l_1$, where $v$ is the particle speed. He found that
for $D \gg v_{\rm w} l$, where $v_{\rm w}$ is the wave phase
velocity, particles diffuse in momentum space according to the
equation
\begin{equation}
\frac{\partial f}{\partial t} = \frac{1}{p^2} \frac{\partial
}{\partial p} \left( p^2 D_p \frac{\partial f}{\partial p}\right),
\label{eq:fp0} 
\end{equation} 
where $f$ is the particle distribution function averaged over
the waves, 
and the momentum diffusion coefficient~$D_p$ is given by
\begin{equation}
D_p = \frac{2 p^2 u_{\rm rms}^2}{9 D},
\label{eq:dp1} 
\end{equation} 
where $p$ is the particle momentum.
The condition $D\gg v_{\rm w} l$ implies
that particles diffuse a distance~$l$ in a time much shorter
than the wave period at scale~$l$.  In this limit, the largest-scale
waves make the dominant contribution to~$D_p$.

This paper extends Ptuskin's analysis to strong compressible MHD
turbulence taking into account the reversible motion of particles
along magnetic field lines.  It is assumed that the rms turbulent
velocity $u_{\rm rms}$ equals or exceeds the Alfv\'en speed~$v_{\rm
A}$, and that there is a single dominant length scale (outer scale)
denoted~$l$ for both velocity and magnetic fluctuations.  Only the
fluctuations at scale~$\sim l$ are considered. As in Ptuskin's (1988)
study, it is assumed that some type of small-scale wave scatters
particles, and that $\lambda_{\rm mfp} \ll l$. It is assumed that
eddies are randomized in a time~$\tau_{\rm rand} \sim l/u_{\rm rms}$,
and that $D_\parallel \gg u_{\rm rms} l$, where~$D_\parallel$ is
the diffusion coefficient for particle motion along field lines. Thus, a particle diffuses a
distance greater than~$l$ during a time~$\tau_{\rm rand}$, and can
diffuse through and then return to an eddy multiple times before the
eddy is randomized in the turbulent flow. It is shown in
section~\ref{sec:dp} that this increases the coherence of stochastic
particle acceleration, effectively increasing~$D_p$ above the value in
equation~(\ref{eq:dp1}) by a factor of $\sim N$, where $N \sim
\sqrt{D_\parallel/u l}$ is the typical number of times a particle
diffuses through and then returns to an eddy before the eddy is
randomized, assuming a particle is tied to a single field line. The
discussion of section~\ref{sec:dp} consists of estimates based on
physical arguments; these estimates are tested and validated by Monte
Carlo particle simulations in section~\ref{sec:MC}.  In
section~\ref{sec:num} the results of the paper are applied to particle
acceleration in the interstellar medium and the lobes of radio
galaxies.

It is known that small cross-field motions and the divergence
of neighboring magnetic field lines in turbulent plasmas allow a
particle to escape from its initial field line by traveling
sufficiently far along the magnetic field.  This places an upper limit
on~$N$, which is explained in the Appendix. It is shown
that this upper limit is not important in the astrophysical
examples considered in section~\ref{sec:num} given the assumed
values of~$D_\parallel$. However, for sufficiently large values of~$D_\parallel$,
the upper limit on~$N$ becomes important.

\section{Phenomenological description of momentum diffusion}
\label{sec:dp}

Since the scattering mean free path is assumed to be~$ \ll l$ and only those eddies
at scale~$\sim l$ are considered, a particle's momentum~$p$ evolves according
to the equation
\begin{equation}
\frac{dy}{dt} = - \frac{\nabla \cdot {\bf u} }{3},
\label{eq:dery} 
\end{equation}
where
\begin{equation}
y = \ln (p/p_0),
\label{eq:defy} 
\end{equation}
$p_0$ is a particle's initial momentum, and
${\bf u}$ is the turbulent velocity associated
with the large-scale eddies (Skilling 1975,
Ptuskin 1988). As a particle diffuses in space, it encounters a series of eddies that
cause random increments in~$y$, leading to diffusion in $y$-space
with diffusion coefficient
\begin{equation}
D_y = \frac{ (\delta y)_{\rm rms}^2}{2\delta t},
\label{eq:Dy} 
\end{equation} 
where $\delta t$ is the time during which increments to~$y$ remain
correlated, and $(\delta y)_{\rm rms}$ is the rms change in~$y$
during a time~$\delta t$.  It is assumed that the turbulent
velocities are randomized in a time
\begin{equation}
\tau_{\rm rand} = \frac{b l }{u_{\rm rms}},
\label{eq:defb} 
\end{equation} 
where $b$ is a constant of order unity.
The characteristic time for a particle to diffuse through an eddy~is 
\begin{equation} 
\tau_{\rm diff} = \frac{l^2}{2 D_\parallel}.
\label{eq:deftaud} 
\end{equation} 
It is assumed that $D_\parallel \gg u_{\rm rms} l$, and thus
\begin{equation}
\tau_{\rm diff} \ll \tau_{\rm rand}.
\label{eq:taucomp} 
\end{equation} 
In this section, it is assumed that a particle remains tied to the
same magnetic field line.  Thus, given equation~(\ref{eq:taucomp}), a
particle typically diffuses through an eddy and later returns to
the same eddy before the eddy is randomized.  During such successive
encounters, the values of~$\nabla \cdot {\bf u} $ within the eddy are
correlated, and  the corresponding increments to~$y$ are
correlated. The correlation time~$\delta t$ for random steps
in~$y$-space is thus~$\tau_{\rm rand}$, and not~$\tau_{\rm diff}$.

The increment to~$y$ during a time~$\tau_{\rm rand}$ is
\begin{equation}
\delta y = - \frac{1}{3} \int_0^{\tau_{\rm rand}} \nabla \cdot {\bf u} dt.
\label{eq:triangley} 
\end{equation} 
It is assumed that there is no average compression or decompression in
the turbulent flow, so that
\begin{equation}
\langle\delta y \rangle = 0,
\label{eq:1} 
\end{equation} 
where $\langle \dots \rangle$ indicates an average over
an arbitrarily large number of random steps.
During a time~$\tau_{\rm rand}$, a particle's position is
distributed over an interval of length~$\sim Nl$, where
\begin{equation}
N = \frac{\sqrt{D_\parallel \tau_{\rm rand}}}{l}.
\label{eq:defN} 
\end{equation} 
Since the values of $\nabla \cdot {\bf u} $ in different
eddies are uncorrelated, the  value of $\nabla \cdot {\bf u}$
averaged over a distance~$Nl$ 
is~$\sim N^{-1/2} (\nabla \cdot {\bf u} )_{\rm rms}$,
where $(\nabla \cdot {\bf u} )_{\rm rms}$ is the
rms value of $\nabla \cdot {\bf u}$. Thus, the rms value
of $\delta y $ is approximately 
\begin{equation}
( \delta y) _{\rm rms} \sim
\frac{N^{-1/2} ( \nabla \cdot {\bf u} )_{\rm rms} \tau_{\rm rand}}{3}.
\end{equation} 
Since only eddies of scale~$l$ are considered, and since $u_{\rm rms}\gtrsim v_{\rm A}$,
\begin{equation}
( \nabla \cdot {\bf u} )_{\rm rms} = \frac{a u_{\rm rms}}{l},
\label{eq:defa} 
\end{equation} 
where $a$ is a constant of order unity.
Thus, if $N\gg 1$, then $( \delta y ) _{\rm rms} \ll 1$, and
the probability $P(y,t)dy$  that $\ln (p/p_0) \in (y, y+dy)$ at time~$t$
satisfies the Fokker-Planck equation
\begin{equation}
\frac{\partial P}{\partial t}
= \frac{\partial^2}{\partial y^2} (D_y P),
\label{eq:fp} 
\end{equation} 
where $D_y$, which is defined in equation~(\ref{eq:Dy}), is given by 
\begin{equation}
D_y \sim \frac{N a^2  u_{\rm rms}^2}{18 D_\parallel}.
\label{eq:dy} 
\end{equation}

The log-momentum diffusion coefficient~$D_y$ is analogous to Ptuskin's
(1988) momentum diffusion coefficient~$D_{p, {\rm Ptuskin}}$ divided
by~$p^2$.  If Ptuskin's coefficient~$D$ for isotropic spatial
diffusion is set equal to~$D_\parallel$,~then in the large-$N$ limit
\begin{equation}
\frac{p^2 D_y}{D_{p, {\rm Ptuskin}}} \sim N.
\label{eq:ration} 
\end{equation}
This ratio can be understood as follows. Ptuskin's result is
approximately recovered by taking $\delta t \sim \tau_{\rm
diff}$. Such a relation is valid if spatial diffusion is isotropic,
since there is only a small chance that a particle returns to the same
eddy multiple times before the eddy is randomized when~$D \gg l u_{\rm
rms}$.  However, when a particle diffuses along a single field line,
the particle's location during a time~$\tau_{\rm rand} = N^2 \tau_{\rm
diff}$ is distributed among roughly~$N$ eddies, and thus a particle
spends a time~$\sim N \tau_{\rm diff}$ in an eddy before the eddy is
randomized. The particle's random walk in~$y$-space can thus be
thought of consisting of individual steps of duration $N\tau_{\rm
diff}$ and magnitude $\delta y = N \tau_{\rm diff} (\nabla \cdot
{\bf u})_{\rm rms}/3$.  The resulting value of~$D_y$ is a factor
of~$N$ larger than in the isotropic-spatial-diffusion case in which a
particle takes individual random steps in $y$-space of
duration~$\tau_{\rm diff}$ and magnitude $\delta y = \tau_{\rm
diff} (\nabla \cdot {\bf u})_{\rm rms}/3$.  Thus, the reversible
motion of a particle along a field line makes the particle's random walk in
$y$-space more coherent, and its diffusion in $y$-space more rapid,
than in the isotropic-spatial-diffusion case.

One consequence of equation~(\ref{eq:dy}) is that $D_y \rightarrow
\infty$ as $\tau_{\rm rand} \rightarrow \infty$. The reason is that
if $\nabla \cdot {\bf u}$ were constant in time along a field line,
increments to~$y$ would remain correlated forever if a particle were tied to that
field line, and the particle would undergo superdiffusion
in~$y$-space.  This phenomenon can be described approximately as
follows.  During a time~$t$, a particle's position is distributed
over~$\sim M$ eddies, where
\begin{equation}
M = \frac{\sqrt{D_\parallel t}}{l}.
\label{eq:defM} 
\end{equation} 
Since the values of $\nabla \cdot {\bf u}$ within different eddies are
uncorrelated, the typical value of $\nabla \cdot {\bf u} $ averaged
over~$M$ eddies is~$\sim M^{-1/2} (\nabla \cdot {\bf u})_{\rm rms}$.
For large~$M$, the rms value of~$y$ for a particle obeys the equation
\begin{equation}
\frac{dy_{\rm rms}}{dt} \sim \frac{M^{-1/2} (\nabla \cdot {\bf u})_{\rm rms}}{3} \propto t^{-1/4}.
\end{equation} 
Integrating and squaring the result, one obtains
\begin{equation}
y_{\rm rms}^2 \sim \frac{16 l(\nabla \cdot {\bf u})_{\rm rms}^2\;  t^{3/2}}{81 \sqrt{D_\parallel }}.
\label{eq:yrms2} 
\end{equation}

\section{Numerical simulations}
\label{sec:MC}

In this section, particle acceleration by compressible turbulence is
investigated with the use of Monte Carlo particle simulations.
Particles are assumed to follow a single field line. Distance along
the field line is denoted by the coordinate~$x$.  The values of
$\nabla \cdot {\bf u} $ at space-time grid points $(x,t) = (ml, n
\tau_{\rm rand})$ are set to either $+\tau_{\rm rand}^{-1}$ or
$-\tau_{\rm rand}^{-1}$ with equal probability, where $m$ and $n$ are
integers.  The value of $\nabla \cdot {\bf u} $ between grid points is
obtained by linear interpolation in space and time, which gives
$(\nabla \cdot {\bf u} )_{\rm rms}\tau_{\rm rand} = 2/3$, or,
equivalently, $ab=2/3$, where $a$ and~$b$ are defined in
equations~(\ref{eq:defb}) and~(\ref{eq:defa}).  During each time step
of duration~$\delta t$, a particle's $y$-value is incremented by an
amount $-\nabla \cdot {\bf u} \delta t/3$, and there is an equal
probability that its $x$-coordinate is either increased or decreased
by an amount~$\sqrt{2D_\parallel \delta t}$.

\begin{figure*}[h]
\vspace{11cm}
\includegraphics{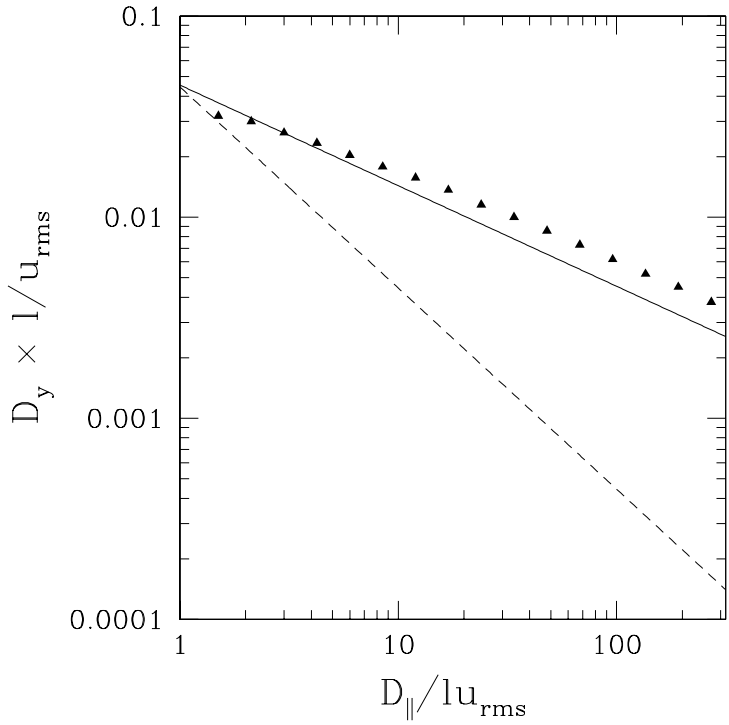}
\caption{$D_y$ as a function of $D_\parallel$. The solid
triangles are the results of the Monte Carlo simulations.
The solid line represents equation~(\ref{eq:dy}), and
the dashed line represents a scaling of $D_y \propto D_\parallel^{-1}$.
\label{fig:f1}}
\end{figure*}

In figure~\ref{fig:f1}, $D_y$ is plotted for different values
of~$D_\parallel$.  To express the results in terms of $u_{\rm rms}$
and~$l$, it is assumed that~$a=1$.  The
values of~$D_y$ in the numerical simulations, depicted with solid
triangles in figure~\ref{fig:f1}, are obtained by dividing the average
of~$y^2$ for $10^4$ simulated particles by $2t$ at $t= 130 \tau_{\rm
rand}$.  The solid line represents the value of~$D_y$ in
equation~(\ref{eq:dy}). The dashed line represents the scaling $D_y
\propto D_\parallel^{-1}$, analogous to Ptuskin's (1988) scaling $D_p
\propto D^{-1}$ for isotropic spatial diffusion of particles.

\begin{figure*}[h]
\vspace{11cm}
\includegraphics{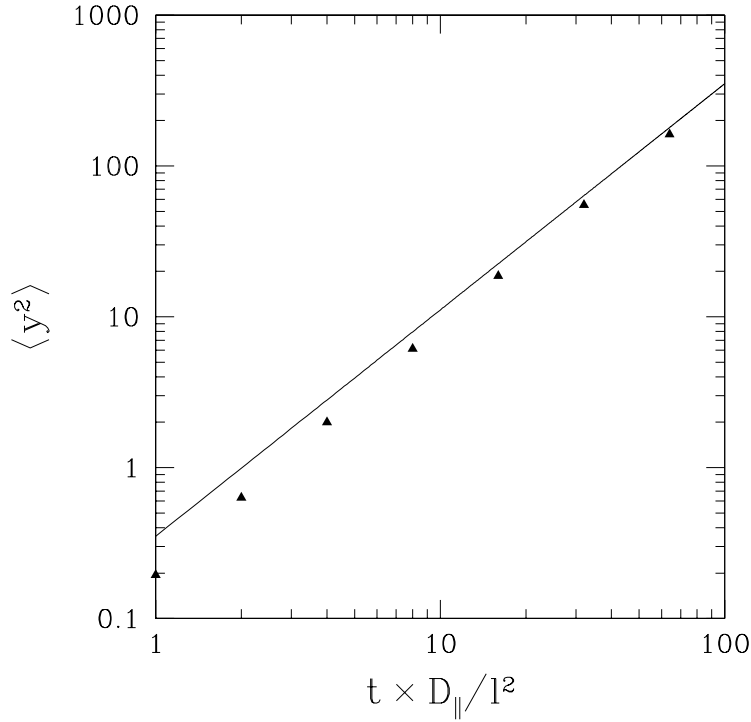}
\caption{Superdiffusion in log-momentum-space when $\nabla \cdot {\bf u}$ is
static along a field line. The solid triangles give the average of $y^2$ for
$10^4$ simulated particles that are tied to a single
magnetic field line. The solid line 
is a plot of equation~(\ref{eq:yrms2}), in which $y_{\rm rms}^2 \propto t^{3/2}$.
\label{fig:f2}}
\end{figure*}

The increase  of $y_{\rm rms}^2$ with time 
when $\nabla \cdot {\bf u}$ is constant in time
along a field line is shown in figure~\ref{fig:f2}.
The solid line is the value of $y_{\rm rms}^2$ in
equation~(\ref{eq:yrms2}).   The
averages of $y^2$ for $10^4$ simulated particles 
are given by the solid triangles and show
superdiffusive behavior in approximate agreement
with  equation~(\ref{eq:yrms2}).

\section{Application to Galactic cosmic rays and relativistic 
electrons in radio galaxies}
\label{sec:num}

From equations~(\ref{eq:fp}) and (\ref{eq:dy}) 
the characteristic time for particle acceleration by large-scale compressible
turbulence is
\begin{equation}
\tau_{\rm acc} = D_y^{-1} \sim \frac{18 D_\parallel^{1/2} l^{1/2}}{u_{\rm rms}^{3/2}},
\label{eq:tauacc} 
\end{equation} 
where it is assumed that $a=1$ in equation~(\ref{eq:defa}) and~$b=1$
in equation~(\ref{eq:defb}).  The values of $D_\parallel$ for Galactic
cosmic rays and relativistic electrons in the lobes of radio galaxies
are not well known. There is disagreement as to the correct way to
model the waves that cause energetic particle scattering (see, e.g.,
Bieber et~al 1994, Schlickeiser \& Miller 1998, Chandran 2000, Felice
\& Kulsrud 2001, Yan \& Lazarian 2002).  Some studies indicate that $D$ and $D_\parallel$
have different values in different phases of the ISM (e.g., Felice \&
Kulsrud 2001), and the values of $D$ and $D_\parallel$ may be
different in the disk and halo of the Galaxy (Berezinskii et~al 1990).
In the hot coronal plasma of the ISM, $u_{\rm rms} \sim 10^7$~cm/s
(Shelton et~al 2001).  This value is comparable to the sound speed,
and thus the turbulence is compressible. It is assumed that $l\sim
3\times 10^{20}$~cm. If $D_\parallel = 10^{28} \mbox{ cm}^2/\mbox{s}$,
then $\tau_{\rm acc} = 3\times 10^7$~yr.  If $D_\parallel = 10^{29} \mbox{
cm}^2/\mbox{s}$, then $\tau_{\rm acc} =  10^8$~yr.  These values
of~$\tau_{\rm acc}$ are comparable to the Galactic confinement time
of~$1$-GeV cosmic rays, and are the types of time scales required if
reacceleration is to explain the  peaking of secondary/primary
ratios at a few~GeV (Ptuskin 2001).  For relativistic electrons in 
radio lobes, it is
simply assumed that $D_\parallel = v_{\rm A} l_{\rm lobe}$, where
$l_{\rm lobe} = 100$~kpc is the typical size of a radio lobe.  For
this value of $D_\parallel$, the time for electrons to diffuse out of
a radio lobe is of order the Alfv\'en crossing time of the lobe. The
thermal plasma density is taken to be $5\times 10^{-5} \mbox{
cm}^{-3}$, which is an an upper limit based upon Faraday
depolarization, and $B$ is taken to be 20~$\mu$G, a typical
equipartition field (Spangler \& Pogge 1984, Spangler \& Sakurai
1985). These values yield~$v_{\rm A}=6\times 10^8$~cm/s, and it is
assumed that $u_{\rm rms} = v_{\rm A}$.
The value of~$l$ for radio-lobe turbulence is taken to be $l_{\rm
lobe}/30$ based on the standard deviation of the linear Stokes
parameters in 3C~166~(Spangler 1983). For these parameters, $\tau_{\rm
acc} \sim 5 \times 10^7$~yr.  This type of time scale is comparable to
the ages of relativistic-electron populations obtained from
synchrotron aging estimates, which suggests that reacceleration may
have to be taken into account in models of radio-lobe electrons.

\section{Summary}
\label{sec:conc} 

It is shown that the reversible motion of energetic particles along
the magnetic field enhances the rate of particle acceleration by
large-scale compressible turbulence relative to the
isotropic-spatial-diffusion case. When this effect is taken into
account, the time scale for acceleration of relativistic electrons by
large-scale compressible turbulence in the lobes of radio galaxies is
$\sim 5\times 10^7$~yr, and the time scale for acceleration of 1-GeV
protons by large-scale compressible turbulence in the coronal plasma
of the ISM is ~$\sim 3-10\times 10^7$~yr. The main source of
uncertainty in these estimates is the uncertainty in~$D_\parallel$;
other sources include the uncertainty in $l$, ~$u_{\rm rms}$, and the
dimensionless constants $a$ and $b$ that relate $\tau_{\rm rand}$ and
$(\nabla \cdot {\bf u})_{\rm rms}$ to $u_{\rm rms}$ and~$l$.

I thank Eliot Quataert, Steve Spangler, Torsten Ensslin, and Eric
Blackman for helpful comments. This work was supported by NSF grant
AST-0098086 and DOE grants DE-FG02-01ER54658 and DE-FC02-01ER54651.

\appendix

\section{Particle escape from field lines through cross-field motion and
field-line divergence}
\label{ap:lrr} 

\begin{figure*}[h]
\vspace{11cm}
\includegraphics{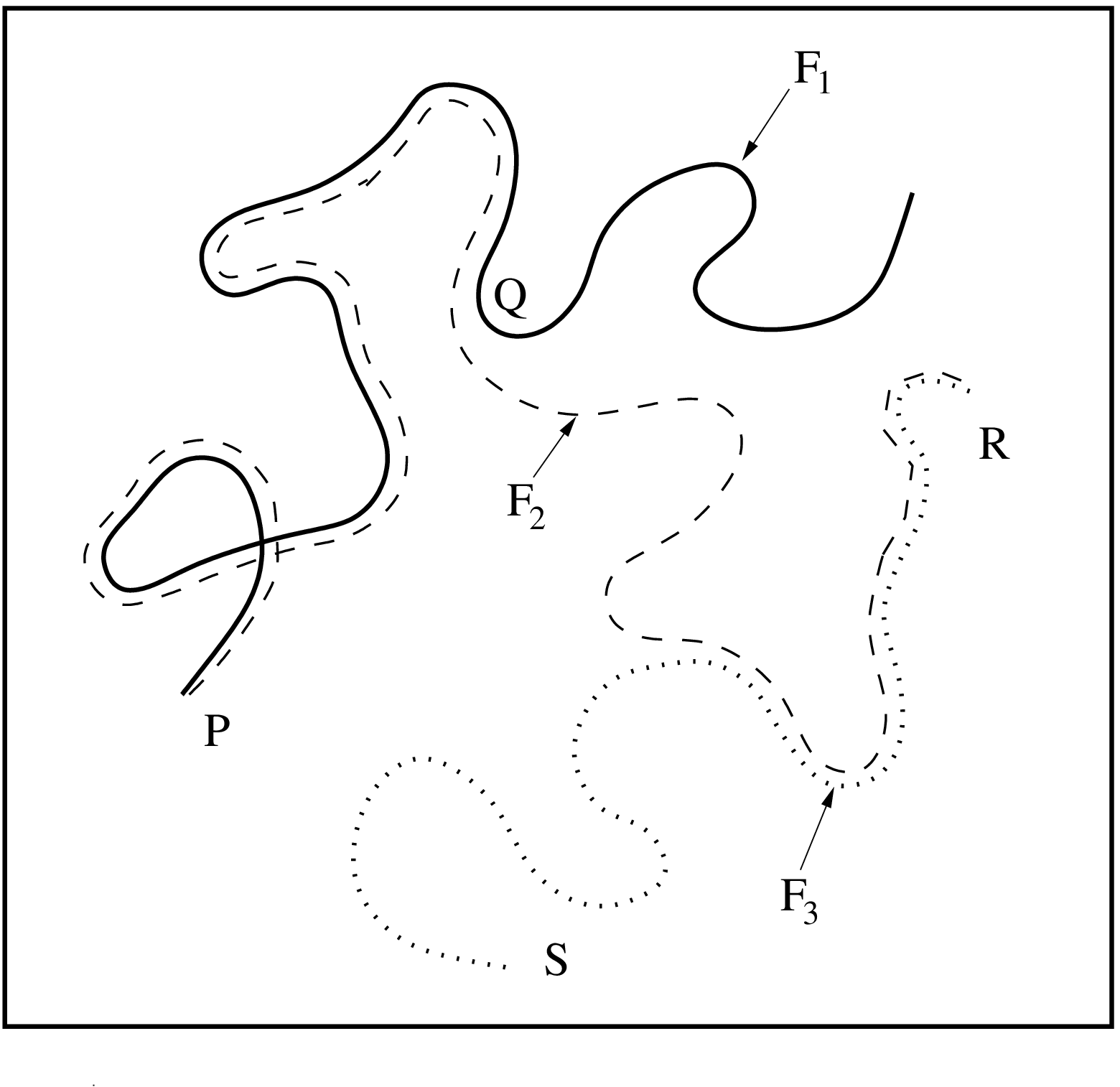}
\caption{Diagram showing the way in which the divergence of
neighboring field lines allows a particle to escape from its
initial field line.
\label{fig:f3}}
\end{figure*}

Particles with gyroradii~$\rho_g$ much less than the dominant
magnetic field scale length~$l$ drift slowly across magnetic field lines
because of  gradients in the magnetic-field strength, curvature
of field lines, and wave pitch-angle scattering. Cross-field motions
enhanced by the divergence of neighboring field lines allow particles
to ``escape'' from their initial field lines, that is, to
separate from their initial field lines by a distance~$l$
(Rechester \& Rosenbluth 1978, Chuvilgin \& Ptuskin 1996, Chandran \& Cowley 1998).
This can be seen with the aid of
figure~\ref{fig:f3}.
Suppose a particle starts out at point~P on field line~F$_1$, and that
its diffusive motion along the field initially moves it towards
point~Q. After moving a short distance, field gradients and collisions
cause the particle to take a step of length~$\sim\rho_g$ across the
magnetic field, from~F$_1$ to a new field line~F$_2$. Although the
particle continuously drifts across the field to new field lines, let
us assume for the moment that it remains attached to~F$_2$.  As the
particle diffuses along~F$_2$, F$_2$ diverges from F$_1$. Let $z_{\rm
s}$ be the distance that F$_2$ must be followed before F$_2$ separates
from F$_1$ by a distance~$l$. (Because the particle continuously
drifts across the field, it typically separates from F$_1$ after
traveling a distance somewhat less than~$z_{\rm s}$ along the field;
this effect is ignored in this discussion.) After the particle moves a
distance~$z_{\rm s}$ along F$_2$, its subsequent motion is not
correlated with~$F_1$. The particle proceeds to point~R, and then its
diffusive motion along the magnetic field changes direction, bringing
it back towards point~Q. Instead of following F$_2$ back to point~Q,
however, the particle drifts across the field and ends up on a new
field line~F$_3$. After following F$_3$ for a distance~$\sim z_{\rm
s}$, the particle separates from F$_2$ by a distance~$\sim l$ and
proceeds to point~S.  In this example, after the particle moves
significantly farther than~$\sim z_{\rm s}$ along the magnetic field,
it has a negligible probability of returning to within a
distance~$\sim l$ of its initial location.  The maximum time a
particle can remain correlated with an eddy is thus of order
\begin{equation}
\tau_{\rm esc} =\frac{\langle z_{\rm s}\rangle^2}{D_\parallel},
\label{eq:tauesc} 
\end{equation} 
where $\langle z _{\rm s} \rangle$ is the average value of $z_{\rm s}$
over a large sample of field-line pairs.

Several studies have calculated $\langle z_{\rm s}\rangle$.  Jokipii (1973) and
Skilling, McIvor, \& Holmes (1974) found that $\langle z_{\rm s} \rangle \sim l$ for
isotropic MHD turbulence with $\delta B \gtrsim B_0$, where $\delta B$
is the fluctuating magnetic field and $B_0$ is the strength of any
background field in the system that is coherent over distances $\gg
l$. Narayan \& Medvedev (2001) found $\langle z_{\rm s} \rangle\sim l$ for
locally anisotropic MHD turbulence with $\delta B\sim B_0$.  A related
quantity, the Kolmogorov entropy, which describes the exponential
divergence of neighboring field lines while their separation is
smaller than the dissipation scale of the turbulence, has been
calculated by several authors (e.g., Rechester \& Rosenbluth~1979,
Zimbardo et~al 1995, Barghouty \& Jokipii 1996, Casse, Lemoine, \&
Pelletier 2001). Another related quantity, the diffusion coefficient
for the wandering of single magnetic field line, has been studied by a
number of authors (e.g., Matthaeus et~al 1995, Michalek \& Ostrowski
1998).

Maron, Chandran, \& Blackman (2003) and Chandran \& Maron (2003) calculated
$\langle z_{\rm s}\rangle$ by tracking field lines in direct numerical
simulations of incompressible MHD turbulence with an inertial
range extending from a dominant length scale~$l$ to a much smaller
dissipation scale~$l_d$. In simulations with $B_0 = 0$, they found that
$\langle z_{\rm s} \rangle$ asymptotes to a value of order
several~$l$ as $\rho_g$ is decreased towards the dissipation scale~$l_d$
in the large-$l/l_d$ limit.
From figures~5 and~7 of
Chandran \& Maron (2003), this value is~$\sim 5-7l$.
We assume that $\langle z_s\rangle$ has a similar value in
turbulence with a mean magnetic field provided~$\delta B \gtrsim B_0$.

The escape of particles from field lines modifies the estimate of~$D_y$
presented in section~\ref{sec:dp}  when $\sqrt{D_\parallel \tau_{\rm rand} } > 
\langle z_{\rm s} \rangle$.  In this case, a particle remains correlated
with a magnetic field line for a time $\tau_{\rm esc}$ that is less
than $\tau_{\rm rand}$. During the time $\tau_{\rm esc}$ the particle travels a
distance $\sim \langle z_{\rm s} \rangle$, and its
location is spread out over roughly 
\begin{equation}
N^\prime = \frac{\langle z_{\rm s}\rangle}{l}
\label{eq:np} 
\end{equation} 
eddies. The arguments leading to equations~(\ref{eq:fp}) and (\ref{eq:dy}) 
can be repeated, replacing $\tau_{\rm rand}$ with $\tau_{\rm esc}$,
to obtain
\begin{equation}
D_y \sim \frac{N^\prime a^2  u_{\rm rms}^2}{18 D_\parallel}.
\label{eq:dyp} 
\end{equation} 
To summarize, if $\tau_{\rm esc} < \tau_{\rm rand}$, then
equation~(\ref{eq:dyp}) is valid. If $\tau_{\rm esc} > \tau_{\rm
rand}$, then equation~(\ref{eq:dy}) is valid. 

For the astrophysical examples considered in section~\ref{sec:num},
$\tau_{\rm esc} \geq \tau_{\rm rand}$, and thus equation~(\ref{eq:dy})   
can be applied. However, for larger values of~$D_\parallel$,
$\tau_{\rm esc}$ can be reduced below~$\tau_{\rm rand}$,
in which case cross-field motion becomes important and
equation~(\ref{eq:dyp}) applies.

\references

Achterberg, A. 1981, A \& A, 97, 259

Barghouty, A., \& Jokipii, J. 1996, ApJ, 470, 858

Berezinskii, V.S., Bulanov, S. V., Dogiel, V. A., Ginzburg, V. L., \&
Ptuskin, V. S. 1990, Astrophysics of Cosmic Rays, (New York:
North-Holland)

Bieber, J., Matthaeus, W., Smith, C., Wanner, W., Kallenrode, M., \& 
Wibberenz, G. 1994, ApJ, 420, 294

Blackman, E. 1999, MNRAS, 302, 723

Brunetti, G., Feretti, L., Giovannini, G., \& Setti, G. 1999, in 
{\em Diffuse thermal and relativistic plasma in galaxy clusters}, eds. H Bohringer,
L. Feretti, P. Schuecker, (Garching:Max-Planck-Institut fur Extraterrestrische Physik),
p. 263

Casse, F., Lemoine, M., \& Pelletier, G. 2001, Phys. Rev., 65, 023002

Chandran, B., 2000, Phys. Rev. Lett., 85, 4656

Chandran, B., \& Cowley, S. 1998, Phys. Rev. Lett., 80, 3077

Chandran, B., \&  Maron, J. 2003, astro-ph/0303214

Cho, J., \& Lazarian, A. 2002, Phys. Rev. Lett., 88, 245001

Chuvilgin, L., \& Ptuskin, V. 1996, A\& A, 279, 278

Dogiel, V. 1999, in
{\em Diffuse thermal and relativistic plasma in galaxy clusters}, eds. H Bohringer,
L. Feretti, P. Schuecker, (Garching:Max-Planck-Institut fur Extraterrestrische Physik),
p. 259

Eichler, D. 1979, ApJ, 229, 409

Eilek, J., \& Henriksen, R.  1984, ApJ, 277, 820

Eilek, J., \& Weatherall, J. 1999, in
{\em Diffuse thermal and relativistic plasma in galaxy clusters}, eds. H Bohringer,
L. Feretti, P. Schuecker, (Garching:Max-Planck-Institut fur Extraterrestrische Physik),
 p.249

Weatherall, J., \& Eilek, J. 1999, in
{\em Diffuse thermal and relativistic plasma in galaxy clusters}, eds. H Bohringer,
L. Feretti, P. Schuecker, (Garching:Max-Planck-Institut fur Extraterrestrische Physik),
p. 255

Felice, G., \& Kulsrud, R. 2001, ApJ, 553, 198

Fermi, E. 1949, Phys. Rev., 75, 1169

Fujita, Y., Takizawa, M., \& Sarazin, C. 2003, ApJ, 584, 190

Gruzinov, A., \& Quataert, E. 1999, ApJ, 520, 849

Hall, D., \& Sturrock, P. 1967, ApJ, 10, 2620

Jokipii, J. 1973, ApJ, 183, 1029

Kulsrud, R.,  \& Ferrari, A. 1971, Astrophys. Sp. Sci., 12, 302

Lithwick, Y., \& Goldreich, P. 2001, ApJ, 562, 279

Manolakou, K., Anastasiadis, A., Vlahos, L. 1999,
A \& A, 345, 653

Maron, J., \& Chandran, B. 2003, astro-ph/0303217 

LaRosa, T., Moore, R., \& Shore, S. 1994, ApJ, 425, 856

Matthaeus, W., Gray, P., Pontius, D., \& Bieber, J. 1995, Phys. Rev. Lett., 75, 2136

Michalek, G., \& Ostrowski, M. 1998, A\&A, 337, 558

Miller, J. 1998, Sp. Sci. Rev., 86, 79

Miller, J., Cargill, P., Emslie, A., Homan, G., Dennis, B., LaRosa, T.,
Winglee, R., Benka, S., \& Tsuneta, S. 1997, J. Geophys. Res., 102, 14631

Miller, J.,  LaRosa, T., \& Moore, R. 1996, ApJ, 461, 445

Narayan, R., \& Medvedev, M. 2001, ApJ, 562, 129

Ohno, H., Takizawa, M., Shibata, S. 2002, ApJ, 577, 658

Ptuskin, V. 1988, Sov. Astron. Lett., 14, 255

Ptuskin, V. 2001, Sp. Sci. Rev. 99, 281

Rechester, A., \&  Rosenbluth, M. 1978, Phys. Rev. Lett., 40, 38

Rechester, A., \&  Rosenbluth, M. 1979, Phys. Rev. Lett., 42, 1247

Schlickeiser, R., \& Miller, J. 1998, ApJ, 492, 352

Shelton, R. et~al 2001, ApJ 560, 730

Skilling, J. 1975, MNRAS, 172, 557

Skilling, J., McIvor, I., \& Holmes, J. 1974, MNRAS, 167, 87P

Spangler, S., \& J. Pogge 1984, ApJ, 89, 342

Spangler, S., \&  Sakurai, T. 1985, ApJ, 297, 84

Spangler, S. 1983, ApJ, 271, L49

Wentzel, F. 1968, ApJ, 152, 987

Yan, H., \& Lazarian, A. 2002, Phys. Rev. Lett., 89, 281102

Zimbardo, G., Veltri, P., Basile, G., \& Principato, S. 1995, Phys. Plasmas,
2, 2653

\end{document}